\documentclass[%
onecolumn,
superscriptaddress,
 amsmath,amssymb,
 aps,
]{revtex4}

\usepackage{graphicx}
\usepackage{dcolumn}
\usepackage{bm}
\usepackage{xcolor}

\begin{document}

\preprint{APS/123-QED}

\title{The generalized Clapeyron equation and its application to confined ice growth}

\author{Robert W. Style}
\email{robert.style@mat.ethz.ch}
\affiliation{%
 Department of Materials, ETH Z\"{u}rich, 8093 Z\"{u}rich, Switzerland.
}%
\author{Dominic Gerber}%
\affiliation{%
 Department of Materials, ETH Z\"{u}rich, 8093 Z\"{u}rich, Switzerland.
}%


\author{Alan W. Rempel}
\affiliation{
 Department of Earth Sciences, University of Oregon, Eugene, Oregon, USA.
}%
\author{Eric R. Dufresne}
\affiliation{%
 Department of Materials, ETH Z\"{u}rich, 8093 Z\"{u}rich, Switzerland.
}%


\date{\today}

\begin{abstract}
Most theoretical descriptions of stresses induced by freezing are rooted in the (generalized) Clapeyron equation, which predicts the pressure that a solid can exert as it cools below its melting temperature.
This equation is central for topics ranging  beyond glaciology to  geomorphology, civil engineering, food storage, and cryopreservation.
However, it has inherent limitations, requiring isotropic  solid stresses  and  conditions  near bulk equilibrium. 
Here, we examine when the Clapeyron equation is applicable by providing a rigorous derivation that details all  assumptions.
We demonstrate the natural extension for anisotropic stress states, and we show how  the temperature and pressure ranges for validity depend on well-defined material properties. 
Finally, we demonstrate how the range of applicability of the (linear) Clapeyron equation can be extended by adding higher-order terms, yielding results that are in good agreement with experimental data for the pressure melting of ice.
\end{abstract}

\maketitle

\begin{figure}[htbp]
    \centering
    \includegraphics[width=9cm]{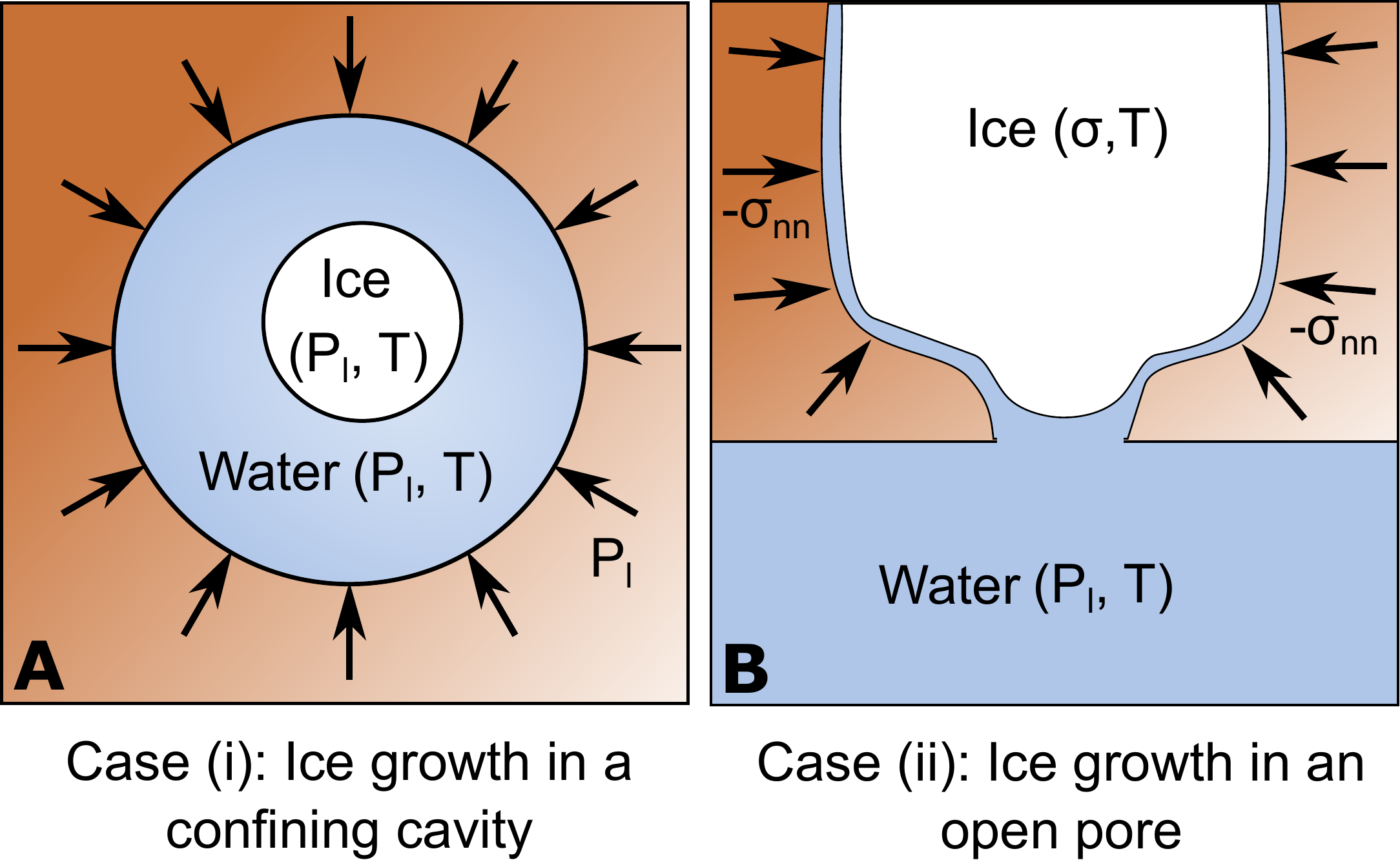}
    \caption{(i) Ice generates pressure as it grows  in a closed cavity, due to the expansion of water upon freezing. (ii) Ice growing in an open pore is fed by nearby water, and this growth wedges open the cavity, generating stresses.}
    \label{fig:fig1}
\end{figure}

When water freezes in confined spaces, it can generate large stresses, often resulting in material damage.
This is important across fields ranging from glaciology to geomorphology, food science, civil engineering and cryopreservation~\cite{dash_physics_2006, rempel_frost_2010, petzold_ice_2009,jha_assessment_2019, karlsson_longterm_1996,vlahou2015freeze,walder1985theoretical}.
Broadly speaking, ice can generate stresses via two different mechanisms \cite{wettlaufer_premelting_2006,peppin2013physics}.
The first is due to the expansion of water as it freezes:
in a closed cavity, freezing will generate pressure (Figure \ref{fig:fig1}a).
The second is unrelated to the expansion of water and often dominates in porous materials \cite{peppin2013physics}.
Here, ice forms in open pores of a wet material (Figure \ref{fig:fig1}b), but no pressure builds up during the initial ice-formation process (any pressure is relieved by water flow away from the growing ice).
However, after the initial ice formation, unfrozen water is sucked back towards the ice crystals.
When this water freezes onto the existing ice, it causes the ice to wedge open its confining pore.
This \emph{cryosuction} process is aided by the presence of thin, mobile layers of water at the surface of ice (known as premelted films) \cite{slater_surface_2019}.
These allow growth of the ice in all directions. 
In both cases ice will continue to grow, building up pressure, until the pressure reaches a maximum value given by a temperature-dependent stall pressure, $P_{st}$ \cite{peppin2013physics,gerber2022stress}.
$P_{st}$ is very similar to the concept of crystallization pressure, found when confined crystals grow from supersaturated solutions \cite{steiger2005crystal,flatt2002salt,desarnaud2016pressure} and to the concept of condensation pressure, when phase separation occurs in confinement \cite{style2018liquid,fernandez2021putting}.

Theoretical descriptions of these stress-generation mechanisms are rooted in the (generalized) Clapeyron equation, a fundamental equation that describes equilibrium between a solid (ice) at pressure $P_s$, and a reservoir of liquid (water) at a different pressure $P_l$~\cite{black_applications_1995, henry_review_2000-1,wettlaufer_premelting_2006}:
\begin{equation}
    \frac{(P_s-P_0)}{\rho_s}-\frac{(P_l-P_0)}{\rho_l}=\frac{q_m(T_m-T)}{T_m}.
    \label{eqn:clap}
\end{equation}
Here, $\rho_l$ and $\rho_s$ are the densities of water and ice respectively, $q_m$ is the specific latent heat of freezing of ice, and $T_m$ is the melting temperature at a reference pressure, $P_0$ (often taken as atmospheric pressure). 
For the freezing mechanisms described above, this equation can be used to predict $P_{st}$ as a function of the temperature, $T$, as at this point, the ice is in equilibrium with the nearby water.
For case (i) with ice growing in a closed cavity, the ice and water are both at the same pressure ($P_s=P_l=P_{st}$), so
\begin{equation}
    (P_{st}-P_0)\left(\frac{1}{\rho_s}-\frac{1}{\rho_l}\right)=\frac{q_m(T_m-T)}{T_m}.\label{eq:2}
\end{equation}
Using values from Table \ref{tab:table1}, we find that ice can exert pressures of $\sim 11$\,MPa per degree of undercooling ($T_m-T$).

For case (ii), the ice and water need no longer have the same pressure.
If the water reservoir is held at the reference pressure $P_l=P_0$, then $P_{st}=P_s$ and
\begin{equation}
    \frac{(P_{st}-P_0)}{\rho_s}=\frac{q_m(T_m-T)}{T_m}.
\end{equation}
In this case, ice can exert pressures of $\sim 1$\,MPa per degree of undercooling.

\begin{table}
\caption{\label{tab:table1}Ice/water parameter values at atmospheric pressure and 273.15K \cite{hobbs2010ice}.}
\begin{ruledtabular}
\begin{tabular}{lll}
  Density of ice & $\rho_s$ & 917 $\mathrm{kg}\,\mathrm{m}^{-3}$\\
  Density of water & $\rho_l$ & 997 $\mathrm{kg}\,\mathrm{m}^{-3}$\\
  Latent heat of fusion & $q_m$ & 334 $ \mathrm{kJ}\,\mathrm{kg}^{-1}$\\
  Melting temperature & $T_m$ & 273.15 K \\
  Heat capacity of ice & $c^p_s$ & 2093 $\mathrm{J}\,(\mathrm{kg\,K})^{-1}$\\
  Heat capacity of water & $c^p_l$ & 4184 $\mathrm{J}\,(\mathrm{kg\, K})^{-1}$\\
  Bulk modulus of ice & $K_s$ & 11.33 GPa\\
  Bulk modulus of water & $K_l$ & 1.96 GPa\\
  Coeff. thermal expansion, ice & $\alpha_s$ & $51\times 10^{-6}$ $\mathrm{K}^{-1}$.\\
  Coeff. thermal expansion, water & $\alpha_l$ & $-50\times 10^{-6}$ $\mathrm{K}^{-1}$.
\end{tabular}
\end{ruledtabular}
\end{table}

Even when ice is not in equilibrium (e.g. it is growing), the Clapeyron equation gives us useful information.
During growth, there is no macroscopic equilibrium, but water immediately adjacent to the ice/water interface can often be considered to be in equilibrium with the ice \cite{wettlaufer_premelting_2006}.
Then, the Clapeyron equation relates the local hydrodynamic pressure in the water, $P_l$, to the local pressure that has been built up in the ice ($P_l$ is just the pressure of a hypothetical, reservoir of water in thermodynamic equilibrium with the ice).
Water flows along nonhydrostatic gradients in $P_l$, so the Clapeyron equation allows us to predict how water is transported towards (or away from) ice, and thus gives ice growth/melting rates \cite{derjaguin_flow_1986,rempel2004premelting,wettlaufer_premelting_2006, wettlaufer_dynamics_1995,style2005surface}.

The various applications of the Clapeyron equation make it  a key tool for understanding freezing processes \citep[e.g.][]{dash_physics_2006,gerber2022stress,vlahou2015freeze,wettlaufer_premelting_2006}. 
However, it makes a number of assumptions.
For example, it assumes that ice can be described by an isotropic pressure, whereas ice is often characterized by an anisotropic stress state, $\sigma_{ij}$ \cite{budd1989review}.
It also uses linear approximations that are  valid only near the bulk melting point of ice (see later).
Thus, several key questions arise. In particular:
What is the appropriate extension of the Clapeyron equation for anisotropically stressed ice?
Over what range of conditions should the Clapeyron equation be applicable?

Surprisingly, we are not aware of a systematic derivation of the Clapeyron equation that would allow us to address these questions.
However, there are  several related works.
For example, several authors have established the thermodynamic relations that govern the dissolution of anisotropically stressed solids into adjacent fluids \citep{gibbs1879equilibrium,kamb1959theory,kamb1961thermodynamic}, with notable  applications to recrystallization  and  pressure solution processes \citep[e.g.][]{paterson1973nonhydrostatic}.
Although melting was not a focus of these works, some of the consequences for ice melting were recognized by \citet{nye1967theory}.
He argued that the phenomenon of wire regelation requires a generalization of equation (\ref{eq:2}) 
where $P_s$ replaced by the normal stress $-\sigma_{nn}$, and not by the mean of the principal stresses $-\mathrm{Tr}(\sigma)/3$, as had been argued by others.
A further, unflinching critique of alternative incorrect theories is given by \citet{kamb1961thermodynamic}. 
Finally, \citet{sekerka2004solid} examined the special case of  a solid  with  $\sigma_{nn}=-P_l$, to show that anisotropically stressed solids in equilibrium with their melt will recrystallize to form a isotropically-stressed state.

Here, we provide a first-principles derivation of the generalized Clapeyron equation, along similar lines to Paterson~\cite{paterson1973nonhydrostatic}.
We clearly lay out all the underlying assumptions, and present the appropriate extension for the melting behavior of anisotropically stressed ice.

\subsection{Deriving the Generalized Clapeyron Equation}
We consider thermodynamic equilibrium for the two scenarios shown in Figure \ref{fig:fig1}, in both of which the temperature is held fixed at $T<T_m$.
In case (i), water freezes in a closed cavity, so that the ice and and water both have the same pressure, $P_s=P_l$.
In case (ii), ice has frozen in an open cavity, and is in equilibrium with neighbouring bulk water, which has pressure $P_l$.
At the same time, the ice exerts a normal stress $-\sigma_{nn}$ on the walls of the cavity, but negligible shear forces, due to the presence of premelted films which lubricate the ice/cavity interface \cite{gerber2022stress}.
The ice cannot grow through the small, connecting pore throat into the neighboring water due to capillarity (i.e. the Gibbs-Thomson effect \citep{hardy1977grain,schollick2016segregated}).

For each scenario, we establish equilibrium behavior by minimizing the relevant free energy of the ice/water system.
The relevant free energy, $G_\mathrm{sys}$ satisfies $\Delta G_\mathrm{sys}=\Delta U_\mathrm{sys}-T\Delta S_\mathrm{sys}+W$, where $U_\mathrm{sys}$ is the internal energy of the ice/water system, $S_\mathrm{sys}$ is its entropy, and $W$ is the work done by the system on its surroundings.
For case (i),
\begin{equation}
    \Delta G_\mathrm{sys}=0=\Delta U_\mathrm{sys}-T\Delta S_\mathrm{sys} + P_l (\Delta V_s+\Delta V_l),
\end{equation}
for case (ii),
\begin{equation}
    \Delta G_\mathrm{sys}=0=\Delta U_\mathrm{sys}-T\Delta S_\mathrm{sys} -\sigma_{nn}\Delta V_s + P_l \Delta V_l,
    \label{eqn:G_sys}
\end{equation}
where $V_s$ and $V_l$ are the volumes of ice and water, respectively.
The first case is just a specialized version of the second, where $-\sigma_{nn}=P_l$.
Thus, without loss of generality, we can proceed with equation (\ref{eqn:G_sys}), and the result will describe both cases.

We consider a small perturbation in the system in Figure \ref{fig:fig1}b, where a small mass of ice, $\Delta m$, melts and flows into the reservoir.
Thus, the volumes of ice and water change as $\Delta V_s=-v_s\Delta m$, and $\Delta V_l=v_l\Delta m$, where $v_s(\mathbf{\sigma}_{ij},T)$ and $v_l(P_l,T)$ are the specific volumes of the ice and water, respectively.
Then equation (\ref{eqn:G_sys}) becomes
\begin{equation}
    u_l\Delta m - u_s\Delta m -T(s_l-s_s)\Delta m + \sigma_{nn} v_s\Delta m + P_l v_l\Delta m=0.
\end{equation}
Here, $u_s(\mathbf{\sigma}_{ij},T)$ and $u_l(P_l,T)$ are the specific internal energies of the ice and water, respectively, and $s_s(\mathbf{\sigma}_{ij},T)$ and $s_l(P_l,T)$ are the respective specific entropies.
Dividing through by $\Delta m$, we obtain

\begin{equation}
    -(\sigma_{nn} v_s + P_l v_l) = (u_l - u_s)-T(s_l - s_s).
    \label{eqn:dU_total}
\end{equation}

In principle, equation (\ref{eqn:dU_total}) completely describes equilibrium between ice and water. \emph{i.e.} one could use tabulated values of $u,v,$ and $s$ to find $-\sigma_{nn}(P_l,T)$.
However, a more convenient form is found by expressing the equation relative to the pressure and temperature under bulk melting reference conditions, ($P_0,T_m$).
Here, $-\sigma_{nn}=P_l=P_0$ so equation (\ref{eqn:dU_total}) becomes
\begin{equation}
   P_0(v_s^o - v_l^o) = (u_l^o - u_s^o) -T_m(s_l^o - s_s^o).
    \label{eqn:dU_total_Tm}
\end{equation}
The superscript $^o$ indicates reference conditions.
Subtracting equations~\ref{eqn:dU_total} and \ref{eqn:dU_total_Tm}, we find
\begin{equation}
    g_l-g_l^o=g_s-g_s^o,
    \label{eqn:eqm}
\end{equation}
where the specific free energies $g_l(T,P_l)=u_l-Ts_l+P_l v_l$, and $g_s(T,\sigma_{ij})=u_s-Ts_s-\sigma_{nn} v_s$.
These can be Taylor-expanded to obtain the Clapeyron equation (e.g. \cite{dash_physics_2006,hutter2008thermodynamic}):
\begin{equation}
    g_l(T,v)=g_l^o(T_m,P_0)+\left(\frac{\partial g_l}{\partial T}\right)_{P_l}(T-T_m)+\left(\frac{\partial g_l}{\partial P_l}\right)_T(P_l-P_0)
     \label{eqn:g1}
\end{equation}
and
\begin{equation}
    g_s(T,\sigma_{ij})=g_s^o(T_m,P_0)+\left(\frac{\partial g_s}{\partial T}\right)_{\sigma_{ij}}(T-T_m)+\left(\frac{\partial g_s}{\partial \sigma_{ij}}\right)_T(\sigma_{ij}+P_0 \delta_{ij}),
     \label{eqn:g2}
\end{equation}
where $\delta_{ij}$ is the identity matrix.
To evaluate the derivatives, we note that $\Delta g_l=-s_l\Delta T + v_l \Delta P_l.$
Thus, at reference conditions, 
\begin{equation}
    \left(\frac{\partial g_l}{\partial T}\right)_{P_l}=-s_l^o,\quad \left(\frac{\partial g_l}{\partial P_l}\right)_T=v_l^o,
    \label{eqn:g}
\end{equation}
and similarly in the solid at reference conditions $(\sigma_{ij}=-P_0\delta_{ij})$
\begin{equation}
    \left(\frac{\partial g_s}{\partial T}\right)_{\sigma_{ij}}=-s_s^o.
\end{equation}
To calculate the final derivative, we notice that $g_s=(f_s+P_0v_s) -\bar{\sigma}_{nn}v_s$, where $f_s$ is the specific Helmholtz free energy of the solid, and  $\bar{\sigma}_{ij}=\sigma_{ij}+P_0\delta_{ij}$. Here, $(f_s+P_0v_s)/v_s^o$ is the free-energy per unit volume of deformations in an atmosphere at constant pressure, $P_0$, and thus is the elastic energy per volume of ice in the reference state.
As such, we can use linear elasticity to write
\begin{equation}
    g_s=\frac{1}{2}\bar{\sigma}_{ij}\epsilon_{ij}v_s^o - \bar{\sigma}_{nn} v_s^o \left(1+\frac{\mathrm{Tr}(\bar{\sigma})}{3K_s}\right),
    \label{eqn:gs}
\end{equation}
where $K_s$ is now the bulk modulus of the solid. $\epsilon_{ij}$ is the strain of the ice relative to its shape in the reference state ($T_m,P_0$), and satsifies the linear-elastic constitutive relationship:
\begin{equation}
    \epsilon_{ij}=\frac{1}{E_s}\left[(1+\nu_s) \bar{\sigma}_{ij}-\nu_s \delta_{ij}\mathrm{Tr}(\bar{\sigma})  \right],
\end{equation}
where $E_s=3K_s(1-2\nu_s)$ is the Young's modulus of the ice and $\nu_s$ is its Poisson ratio.
For small strains, $v_s=v_s^o(1+\mathrm{Tr}(\epsilon))$, and we use this in the second term of equation (\ref{eqn:gs}).

With the two equations above, we can evalulate the remaining derivative at $ (P_0,T_m)$:
\begin{equation}
    \left(\frac{\partial g_s}{\partial \sigma_{ij}}\right)_T(\bar{\sigma}_{ij}=0)=-n_i n_j v_s^o.
    \label{eqn:anisotropic}
\end{equation}
Here, $n_i$ is the normal vector to the surface of the ice, so that $\bar{\sigma}_{nn}=n_i\bar{\sigma}_{ij}n_j$.

Finally, we can insert these first-derivative expressions into equations (\ref{eqn:eqm}--\ref{eqn:g2}) to obtain the Clapeyron equation for anisotropically stressed solids:
\begin{equation}
    -\frac{(\sigma_{nn}+P_0)}{\rho_s^o} - \frac{(P_l-P_0)}{\rho_l^o} = \frac{q_m(T_m-T)}{T_m}.
    \label{eqn:gce}
\end{equation}
Here, $\rho_l^o=1/v_l^o$, and $\rho_s^o=1/v_s^o$ are the densities of water and ice respectively at the bulk melting point, and $q_m\equiv(s_l^o-s_s^o)T_m$.
Consistent with the regelation analysis of \citet{nye1967theory}, this version of the Clapeyron equation is identical to equation (\ref{eqn:clap}), but with $P_s$ replaced by $-\sigma_{nn}$, and not $-\mathrm{Tr}(\sigma)/3$, as one might naively assume. 

\subsection{Field data supporting the anisotropic Clapeyron equation}

Glaciological field data supporting the form of equation (\ref{eqn:gce}) comes from simultaneous measurements of temperatures and liquid pressures in glacier boreholes.
These measurements show that temperatures increase when changes in the hydrologic system cause borehole pressures, $P_l$, to decrease \citep[e.g.][]{andrews2014direct,huss2007glacier}.

The anisotropic Clapeyron equation indeed recovers this correlation.
Along borehole walls, $\sigma_{nn} = -P_l$.
Inserting this into equation (\ref{eqn:gce}), we find that changes in temperature are correlated with changes in borehole pressure by:
\begin{equation}
    \Delta T =-\frac{T_m}{q_m}\left(\frac{1}{\rho_s^0}-\frac{1}{\rho_l^0}\right)\Delta P_l\approx \left(-7.16\times 10^{-8}\,\mathrm{K\,Pa^{-1}}\right)\Delta P_l\;,\label{eq:bh}
\end{equation}
in agreement with the field data.

By contrast, naively extending the isotropic Clapeyron equation (\ref{eqn:clap}), by replacing $-P_s=\mathrm{Tr}(\sigma)/3$, does not match the experimental data.
The classic analysis of \citet{nye1953flow} gives the complete stress tensor at the surface of an idealized cylindrical borehole containing liquid at pressure $P_l$.
Far from the borehole, the ice has a far-field isotropic ice pressure $P_\infty$, and creeps according to Glen's flow law with exponent $n=3$ \cite{glen1955creep,hewitt2019model}.
In this case, $-\mathrm{Tr}(\sigma)/3=P_l +\left(P_\infty-P_l\right)/n$.
Substituting $P_s=-\mathrm{Tr}(\sigma)/3$ into  the isotropic Clapeyron equation (\ref{eqn:clap}) and treating the far-field ice pressure as constant leads to
\begin{equation}
    \Delta T
    =-\frac{T_m}{q_m}\left(\frac{1}{\rho_s^0}-\frac{1}{\rho_l^0}-\frac{1}{n\rho_s^0}\right)\Delta P_l\approx\left(2.26\times 10^{-7}\,\mathrm{K\,Pa^{-1}}\right)\Delta P_l\;.\label{eq:trresult}
\end{equation}
This predicts the opposite of the correlation seen in the field data.

\subsection{Errors in the Clapeyron equation}
In deriving this version of the Clapeyron equation, we have had to make two main assumptions.
Firstly, strains in the ice are small, so we can use linear elasticity \cite{sekerka2004solid}.
This is reasonable as the stresses in the ice (which are $O$(MPa) -- see introduction) are much less than the ice's elastic moduli $E_s,K_s = O$(GPa), so strains will be small.

Secondly, we assume that higher-order terms in the expansions of $g_l$ and $g_s$ are negligible.
We can test this by reverting to the case of isotropically-stressed ice $(\sigma_{ij}=-P_s\delta_{ij})$.
Then, we Taylor-expand equation (\ref{eqn:eqm}) in $T,P_l$ and $P_s$ to obtain the second-order version of the Clapeyron equation:
\begin{eqnarray}
    \label{eqn:gce2}
    \frac{(P_s-P_0)}{\rho_s^o} - \frac{(P_l-P_0)}{\rho_l^o} =&& \frac{q_m(T_m-T)}{T_m}\\ &&-\frac{c_l^p-c_s^p}{2T_m}(T_m-T)^2-\frac{1}{2\rho_l^o K_l}(P_l-P_0)^2+\frac{1}{2\rho_s^o K_s}(P_s-P_0)^2\nonumber \\
    &&-\frac{\alpha_l}{\rho_l^o}(T_m-T)(P_l-P_0)+\frac{\alpha_s}{\rho_s^o}(T_m-T)(P_s-P_0).\nonumber
\end{eqnarray}
Here, we use the following identities \cite{venerus2018modern}:
\begin{equation}
    \frac{\partial^2 g}{\partial T^2}=-\frac{c^p}{T_m}, \quad
    \frac{\partial^2 g}{\partial P^2}=-\frac{1}{K \rho^o}, \quad 
    \frac{\partial^2 g}{\partial P\partial T}=\frac{\alpha}{\rho^o}.
\end{equation}
$c^p$ is the heat capacity at constant pressure, $K$ is again the isothermal bulk modulus, and $\alpha$ is the coefficient of thermal expansion.

\begin{figure}[htbp]
    \centering
    \includegraphics[width=9cm]{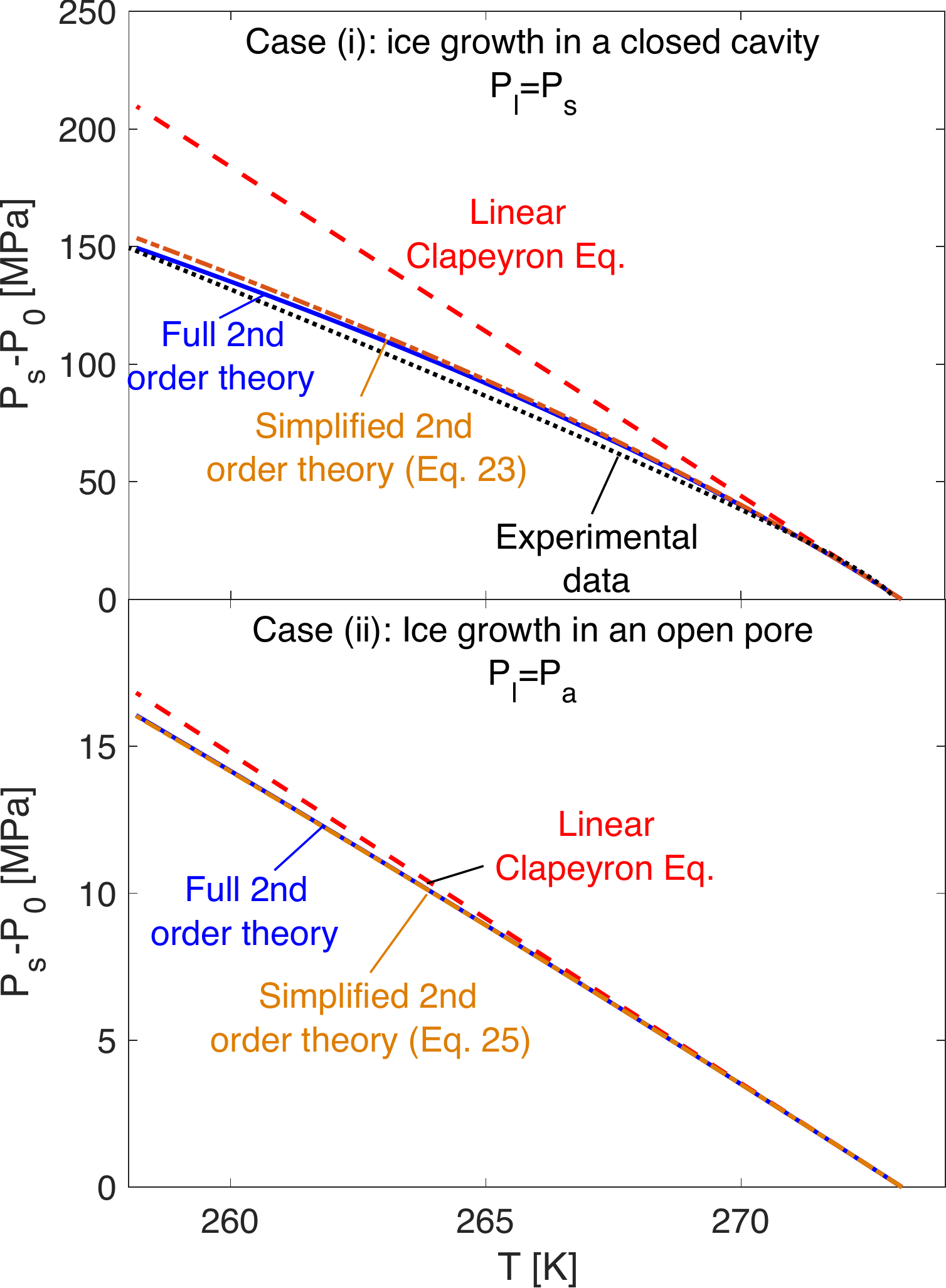}
    \caption{Evaluating the accuracy of the Clapeyron Equation. A) The pressure of ice in bulk ice/water equilibrium in a closed cavity ($P_l=P_s=-\sigma_{nn}$), as a function of undercooling. The black, dotted curve shows experimental data \cite{dunaeva2010phase}. B) The stress exerted by ice in an open pore, as a function of undercooling. Both figures show the linear Clapeyron equation (dashed red), full 2nd-order theory (Eq. \ref{eqn:gce2}, blue), and simplified 2nd-order theory (Eqs. \ref{eq:2nd_order_simp1},\ref{eq:2nd_order_simp2}, orange dash-dotted).} 
    \label{fig:results}
\end{figure}

We can now predict the pressure-melting curve for different freezing scenarios. 
For bulk ice/water equilibrium (Figure \ref{fig:fig1}a), $P_s=P_l$, and we take atmospheric pressure, $P_a$, as the reference pressure.
Figure \ref{fig:results}a compares the isotropic Clapeyron equation (\ref{eqn:clap}) (red, dashed) with experimental data (black, dotted) \cite{dunaeva2010phase}.
There is a significant error between the two results for an undercooling of more than $\sim 3\,^\circ$C.
However, when we use the full, second-order Clapeyron equation (\ref{eqn:gce2}) (blue), we find good agreement down to an undercooling of at least 15$\,^\circ$C.
In this situation, the terms that are quadratic in pressure dominate the error, and to excellent approximation (Figure \ref{fig:fig1}a, orange dash-dotted):
\begin{equation}
    \left(\frac{1}{\rho_s^o}-\frac{1}{\rho_l^o}\right)(P_s-P_0)+\left(\frac{1}{2\rho_l^o K_l}-\frac{1}{2\rho_s^o K_s}\right)(P_s-P_0)^2=\frac{q_m(T_m-T)}{T_m}.
    \label{eq:2nd_order_simp1}
\end{equation}
Comparing the first two terms in this equation, we see that the linear equation (\ref{eqn:gce}) is only appropriate when:
\begin{equation}
    |P_s-P_a|\ll \Delta P^*=\left|\left(\frac{1}{\rho_s^o}-\frac{1}{\rho_l^o}\right)\left(\frac{1}{2\rho_s^oK_s}-\frac{1}{2\rho_l^oK_l}\right)^{-1}\right|\approx 420\mathrm{MPa}.
\end{equation}

We can perform a similar analysis for freezing in an open system (Figure \ref{fig:fig1}b).
We let $P_l=P_0=P_a$, and assume that the ice exerts an isotropic pressure, $P_s$ on its surroundings.
Figure \ref{fig:results}b compares the prediction of the Clapeyron equation (\ref{eqn:clap}) (red, dashed) with that obtained when we keep the extra quadratic terms (\ref{eqn:gce2}) (blue).
We are not aware of any experimental data precise enough to validate the theory \cite{gerber2022stress}.
However, here, the higher-order theory agrees well with the linear Clapeyron equation down to large undercoolings.
The difference is dominated by the term in equation (\ref{eqn:gce2}) that is quadratic in undercooling.
Thus, to excellent approximation (Figure \ref{fig:fig1}a, orange dash-dotted):
\begin{equation}
    P_s-P_a=\frac{\rho_s^o q_m(T_m-T)}{T_m} - \frac{\rho_s^o(c_l^p-c_s^p)}{2T_m}(T_m-T)^2.
    \label{eq:2nd_order_simp2}
\end{equation}
Comparing terms on the right-hand side, shows that we only recover the linear Clapeyron equation (\ref{eqn:clap}) if
\begin{equation}
    |T_m-T|\ll\Delta T^*=\left| \frac{2q_m}{c_l^p-c_s^p} \right|\approx 320 \mathrm{K}.
\end{equation}
This requirement is certainly satisfied for most terrestrial temperatures.
Thus, there is some justification for use of the linear Clapeyron equation down to relatively large undercoolings to model this type of freezing scenario.

To summarize, our results suggest that the linearized Clapeyron equation will be valid, provided that $|P_l-P_0|$ and $|P_s-P_0|$ are both small relative to $\Delta P^*$, while $|T_m-T|\ll \Delta T*$.
At larger pressures/undercoolings, the quadratic terms in equation (\ref{eqn:gce2}) should be included.

\subsection{Conclusions}

In conclusion, we have derived the linear, Clapeyron equation describing equilibrium between water and ice, clearly laying out all the assumptions involved.
In particular, this equation is derived using a Taylor expansion around a reference temperature and pressure, and ignoring higher-order terms.
Thus, it is only valid for a range of pressures and and temperatures around the reference conditions.
Fortunately, for most naturally-occurring terrestrial freezing scenarios, the linear form of the Clapeyron equation should be adequate.
For example, at the base of a glacier, pressures are typically close to hydrostatic, and thus $O($MPa) \cite{sugiyama2004short} -- this is small enough to lie within the range of applicability of the Clapeyron equation.
However, more extreme conditions are expected in extraterrestrial settings \cite[e.g.][]{dunaeva2010phase,mccarthy2016tidal}.
There, the linearized Clapeyron equation will not accurately predict melting temperatures, which could lead to significant errors in models of ice dynamics (as predicted flow rates are typically based on the departure from bulk melting conditions \cite{budd1989review}).
In this case, the accuracy of the Clapeyron equation can be improved by retaining higher-order terms in the Taylor expansion.

We have also demonstrated the correct form of the Clapeyron equation for the case where ice is anisotropically stressed.
This is identical to the isotropic form of the Clapeyron equation, but with ice pressure, $P_s$ replaced by the normal stress exerted by ice on its surroundings, $-\sigma_{nn}$.
One consequence of this, is that differently stressed faces of ice (for example in a polycrystal) will have different melting temperatures.

While our analysis has focused on ice and water, the results should apply to any processes involving solid/liquid equilibrium, for example in the melting and deformation of rocks in geological processes (e.g. \cite{katz2006dynamics}). 
Note however, that there are two, key further effects that will likely be important to include in real-world applications.
Firstly, we have neglected the presence of solutes, which are known to strongly affect the solid/liquid equilibrium \cite{zhang2021quantitative,wettlaufer1999impurity,zhou2018application,dedovets2018five}.
Secondly, we we have ignored the surface energy of the ice \cite{wettlaufer_premelting_2006,wilen1995giant}.
However, we anticipate that both of these effects can be incorporated into the results presented here, by including colligative and capillary effects in the analysis above.

\begin{acknowledgments}
RWS and DG acknowledge support from an ETH Research Grant (Grant No. ETH-38~18-2), and from the Swiss National Science Foundation (Grant No. 200021-212066); AWR received funding from NSF-2012468 and a UO Faculty Research Award.  
\end{acknowledgments}

\end{document}